\documentclass[aip,graphicx,reprint,floatfix]{revtex4-1}

\usepackage{float}
\usepackage{graphicx}
\usepackage{chemformula}
\usepackage{gensymb} 
\usepackage{upgreek}

\newcommand{\ourtable}{~SI~} 

\newcommand{\un}   [1]{\ensuremath{\,\mathrm{#1}}}

\newcommand{\pderl}[2]{\ensuremath{\partial #1/\partial #2}}

\newcommand{\mySiN}{Si\textsubscript{3}N\textsubscript{4}~}

\draft 

\begin{document}


\title{Efficient optomechanical mode-shape mapping of micromechanical devices}



\author{David Hoch}
\affiliation{Department of Physics, Technical University of Munich, Garching, Germany}
\affiliation{Munich Center for Quantum Science and Technology (MCQST), Munich, Germany}
\affiliation{Institute for Advanced Study, Technical University of Munich, Garching, Germany}

\author{Kevin-Jeremy Haas}
\affiliation{Department of Physics, Technical University of Munich, Garching, Germany}

\author{Leopold Moller}
\affiliation{Department of Physics, Technical University of Munich, Garching, Germany}

\author{Timo Sommer}
\affiliation{Department of Physics, Technical University of Munich, Garching, Germany}
\affiliation{Munich Center for Quantum Science and Technology (MCQST), Munich, Germany}

\author{Pedro Soubelet}
\affiliation{Department of Physics, Technical University of Munich, Garching, Germany}
\affiliation{Walter Schottky Institute, Technical University of Munich, Garching, Germany}

\author{Jonathan Finley}
\affiliation{Department of Physics, Technical University of Munich, Garching, Germany}
\affiliation{Walter Schottky Institute, Technical University of Munich, Garching, Germany}

\author{Menno Poot}
\email{menno.poot@tum.de}
\affiliation{Department of Physics, Technical University of Munich, Garching, Germany}
\affiliation{Munich Center for Quantum Science and Technology (MCQST), Munich, Germany}
\affiliation{Institute for Advanced Study, Technical University of Munich, Garching, Germany}

\date{\today}

\begin{abstract}
We demonstrate a method to optically map multiple modes of mechanical structures simultaneously. The fast and robust method, based on a modified phase-lock-loop, is demonstrated on a silicon nitride membrane and compared with three different approaches. Line traces and two-dimensional maps of different modes are acquired. The high quality enables us to determine the weights of individual contributions in superpositions of degenerate modes. 
\end{abstract}


\maketitle 
In recent years, there have been many applications for integrated opto- and electromechanics extending from e.g. mobile communication \cite{hesjedal_APL_AFM_SAW} and highly sensitive sensors \cite{masssensor, naik_natnano_masssensing, mckeown_MNE_gassensor, forcesensor, bleszynski-jayich_science_persistent_currents, fong_nature_casimir, singh_PRL_acoustic_blackbody_detection} to position detection close to the quantum limit \cite{lahaye_science_quantum_limit, etaki_natphys_squid, anetsberger_natphys_nearfield}. In the development of such devices, an efficient method for mode characterization is instrumental and, hence, a number of techniques including optical interferometry \cite{barg_APB_darkfield_mode_imaging, zhang_APL_ultrathin_membranes, davidovikj_NL_graphene_mode_visualization}, heterodyne detection \cite{shen_RSI_image_AlN_microdisk_vibrations, Romero_PRAppl_SiN_phonon_waveguide_heterodyn}, dark field imaging \cite{singh_PRL_acoustic_blackbody_detection, barg_APB_darkfield_mode_imaging}, and force microscopy \cite{hesjedal_APL_AFM_SAW, garcia_NL_imaging_graphene, garcia_PRL_dfm_nanotube, etaki_natphys_squid} have been developed to visualize mechanical modes. However, most of these have one or more drawbacks, such as poor sensitivity, lacking phase information, low spatial resolution, or long measurement times.
%
%
%
Here, we demonstrate an experimental method that combines the high sensitivity of the optical interferometric techniques with demodulation and frequency tracking to offer rapid and robust imaging of multiple modes at the same time. The advantages of the technique are illustrated by mapping the eigenmodes of a square silicon nitride (SiN) membrane. With this method, they can not only be unambiguously identified, but also their mode composition can be determined quantitatively and insights in clamping losses are provided.

\begin{figure}[htbp]
\includegraphics[width=1.0\columnwidth]{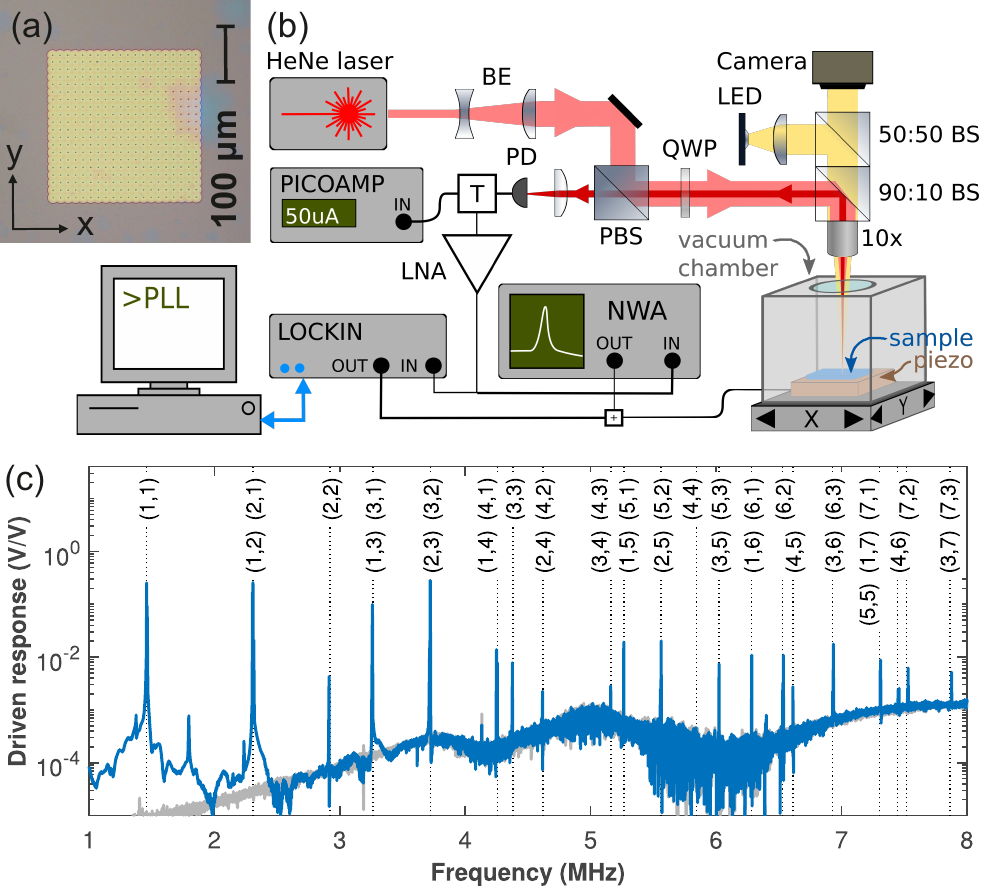}
\caption{
(a) Optical micrograph of the membrane 
(b) Schematic overview of the measurement setup. BE: beam expander, (P)BS: (polarizing) beam splitter, QWP: quarter wave plate, PD: photo detector, LNA: low-noise amplifier, NWA: network analyzer, T: bias tee, +: combiner, LED: light emitting diode for illumination.
(c) Driven response of the membrane measured using the NWA (blue) with the calculated frequencies [Eq. (\ref{eq:f})] as black dashed lines and the mode numbers indicated. The gray trace is the instrument background. 
\label{fig:intro}}
\end{figure}

The SiN membranes are made on chips with 330 nm high-stress \ch{Si3N4} \cite{hoch_TJP_onchip, hoch_Sbeam, terrasanta_AlN_on_SiN}. Release holes are defined using electron-beam lithography followed by a fluorine-based reactive ion etch, exposing the underlying \ch{SiO2} \cite{adiga_APL_SiN_drum_Q_mode}. The membranes are released using buffered hydrofluoric acid followed by critical-point drying; a micrograph of the final suspended membrane is shown in Fig. \ref{fig:intro}(a). The reflectivity of the structure depends on the distance between the membrane and Si substrate, enabling interferometric measurements of the membrane displacement using the setup shown in Fig. \ref{fig:intro}(b). For this, a HeNe laser is focused using a 10x microscope objective with a 32 mm working distance and NA=0.28. It has a fixed position outside the vacuum chamber, which is mounted on a motorized x-y stage to scan with steps of $1.25\un{\mu m}$ while measuring the reflected light using a photo detector. For excitation and detection, either a network analyzer (NWA, HP4396A) or lock-in amplifier (LIA, Zurich Instruments HF2) can be used. Their output goes to the piezo-electric actuator to excite the membrane. Its vibrations modulate the light on the photo detector, which is again detected with the NWA or LIA. The dc reflection can be recorded using a picoamp current meter. For further details see App. \ref{app:demod}.

The out-of-plane modes of a square membrane with side lengths $a$ under uniform tension can be calculated analytically \cite{strauss_PDE}. The normalized mode shapes are:
\begin{equation}
    \xi_{m,n}(x,y) = \sin(\pi m x/a)\sin(\pi n y/a), \label{eq:u}
\end{equation}
so that the local displacement is \cite{poot_physrep_quantum_regime} $u_{m,n}(x,y) = U_{m,n} \xi_{m,n}(x,y)$. The modes are labeled using two integers $m$ and $n$ that count the number of anti-nodes, at which the amplitude is $U_{m,n}$, in the x and y direction, respectively. The $(m,n)$ mode, thus, has $m-1$ ($n-1$) vertical (horizontal) nodal lines. The corresponding eigenfreqencies are:
\begin{equation}
    f_{m,n} =  f_{1,1}\times\left(\frac{m^2+n^2}{2}\right)^{1/2}; ~ f_{1,1} = \frac{1}{2a} \left(\frac{2\sigma}{\rho}\right)^{1/2}. \label{eq:f}
\end{equation}
Here, $\rho = 3.17 \times 10^3 \un{kg/m^3}$ is the mass density of \mySiN \cite{lide_chemphys} and $\sigma \sim 1.05 \un{GPa}$ is the film stress in our wafers \cite{hoch_Sbeam} yielding $f_{1,1} = 1.48 \un{MHz}$ for $a=275 \un{\mu m}$. 

Experimentally, the eigenfrequencies appear as a series of sharp resonances in Fig. \ref{fig:intro}(c). The first peak is at $1.46 \un{MHz}$, close to the result from Eq. \eqref{eq:f} which also shows that once $f_{1,1}$ is known, the other eigenfrequencies can be calculated; their values (dashed lines) nicely match the observed peaks so that resonances can be identified. For example, the peak at 2.92 MHz matches $f_{2,2}$. On the other hand, the one at 3.26 MHz coincides with both (1,3) and (3,1). Theoretically, a perfectly square membrane has degenerate modes, i.e. $f_{m,n} = f_{n,m}$ but, in practice, small imperfections can break the degeneracy. When zooming in, two peaks with $\sim 1 \un{kHz}$ splitting are visible (Fig. \ref{fig:31_and_13}). Still, from their frequencies alone these cannot be identified. Instead, their mode shape should be measured to unambiguously determine which peak corresponds to which mode.

\begin{figure}[tbhp]
  \includegraphics[width=1.0\columnwidth]{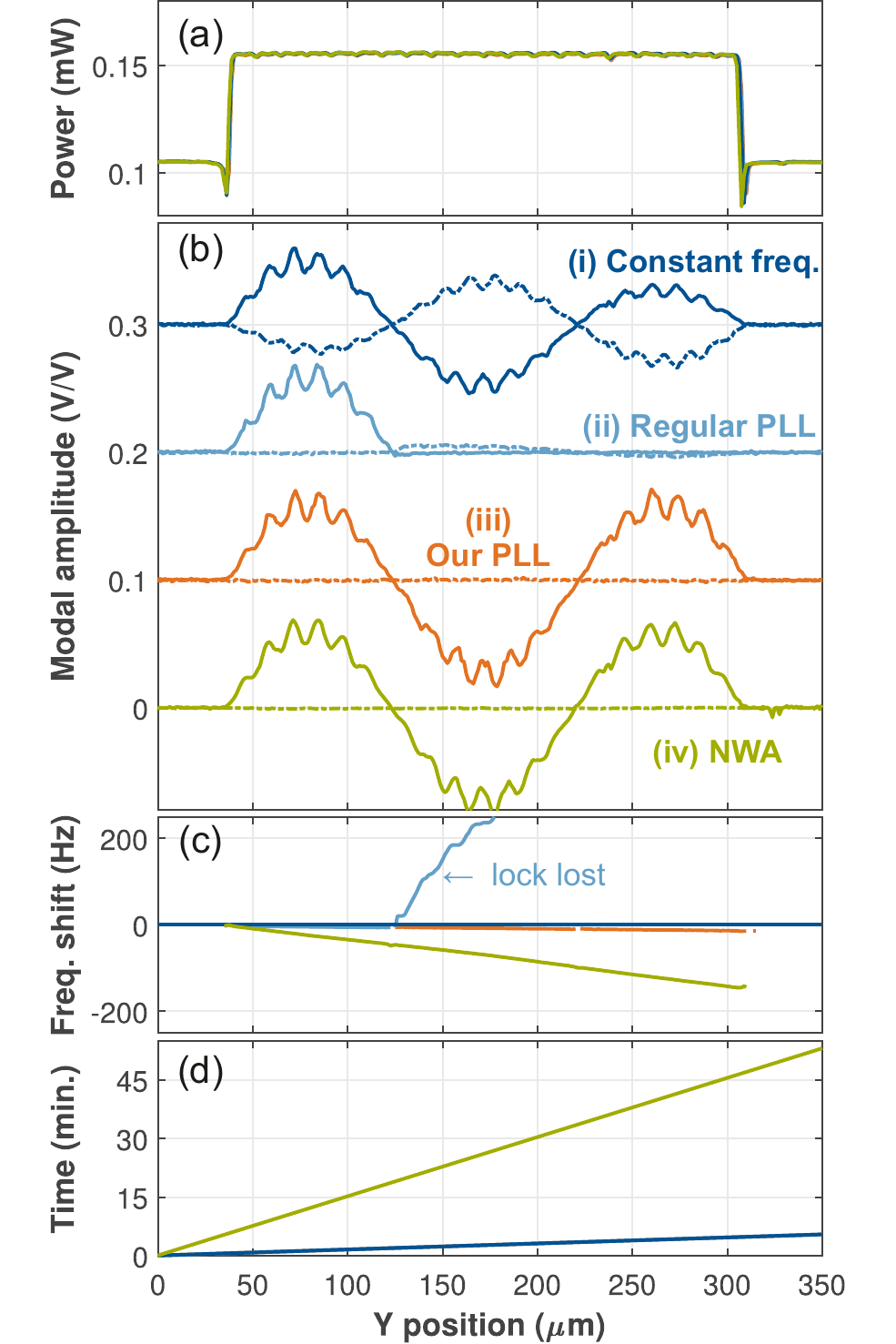}
  \caption{Line traces over the membrane for (i) constant frequency (dark blue), (ii) regular PLL (light blue), (iii) the modified PLL with mod $180\degree$ (orange), (iv) data measured using a network analyzer (green).
  (a) Reflected laser power.
  (b) The mode profile of the (1,3) mode near 3.26 MHz. Solid (dashed) lines indicate the real (imaginary) part. The curves are offset for clarity. 
  (c) Frequency change during the measurement. For (i)-(iii), this was the actuation $f$, whereas for (iv) the resonance $f_0$ was extracted from the NWA traces. By definition the frequency was constant in (i).
  (d) Elapsed time since the start of the trace. \label{fig:traces}}%
\end{figure}

\begin{figure*}[!hbtp] 
\includegraphics[width=1.0\textwidth]{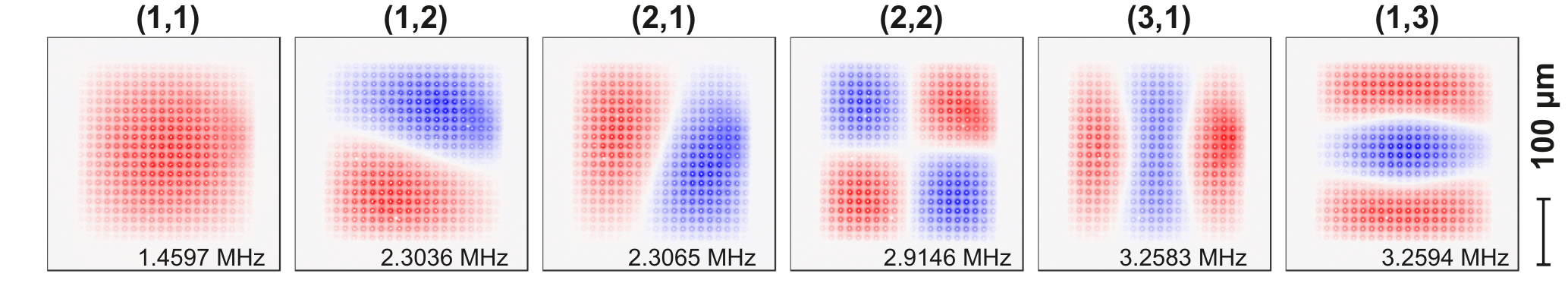}
\caption{
Normalized mode maps (real part) of the first six modes of the square SiN membrane acquired in a single measurement with our PLL method. The release holes are visible as an array of dots across the entire surface of the structure. Starting frequencies are indicated in each map; further properties are listed in Table \ourtable.}
\label{fig:6modes}
\end{figure*}

Before presenting the full two-dimensional mode maps, we first compare line traces taken with different methods in Fig. \ref{fig:traces} to show the robustness and efficiency of ours. The membrane is scanned in the $y$ direction while acquiring the signal of the $3.260 \un{MHz}$ mode using four different methods. First of all, Fig. \ref{fig:traces}(a) shows the dc reflection, which overlaps for all methods. The suspended membrane has a higher reflectivity compared to the supported regions and the holes are visible as small dips in the signal. 

Now, method (i) for obtaining a mode shape (dark blue lines in Fig. \ref{fig:traces}) is to simply drive the mechanical resonator at that resonance frequency and record the amplitude\cite{miller_APL_mode_map_blue_red}.
%
%
%
%
Figure \ref{fig:traces}(b) shows that the suspended part of the membrane has a clear response with small modulations due to the holes. There are two nodes in the modal amplitude vs. $y$, indicating that this is the (1,3) and not the (3,1) mode. Taking a closer look shows that, unlike for the theoretical prediction of Eq. \eqref{eq:u}, the  anti-nodes have unequal magnitudes. Also, the modal amplitude shows an imaginary part (App. \ref{app:demod}) that grows with time, indicating that the resonance frequency drifted from the (fixed) driving frequency during the measurement. This means that the naive approach (i) does not yield accurate mode shapes.


A standard approach to track a resonance is a phase-lock loop (PLL) \cite{rohse_RSI_cavity_lock_optical_spring, bleszynski-jayich_science_persistent_currents}. Here, it is a software-implemented PI-controller in LabVIEW in combination with digital demodulation in the LIA\footnote{Note, that it is also possible to perform this task using the digital signal processor in the lock-in amplifier \cite{poot_PRA_squeezing_feedback,poot_NJP_Yfeedback}, which can further improve the operation speed.}. The PI-controller updates the driving frequency $f$ to keep the phase $\phi$ at the setpoint $\phi_\mathrm{sp}$ using:
\begin{equation}
f_{n+1} = f_1 + P e_n + I \sum_{j=1}^n e_j. \label{eq:PLL}
\end{equation}
Here, $e_n = \phi_n - \phi_\mathrm{sp}$ is the error in the $n$-th sample, and $P$ and $I$ are the proportional and integral gain, respectively. Fig. \ref{fig:traces}(b) shows that with this regular PLL [method (ii), light blue], the first anti node has a lower imaginary part compared to the previous method. However, as also indicated by the sudden large shift in Fig. \ref{fig:traces}(c), the PLL loses lock after the node where the mode changes sign resulting in a $\pi$ jump in $\phi$. This problem motivates our changes to the regular PLL. For method (iii) we added a modulo operation: $e_n \rightarrow e_n \mod \pi$. This way, the PLL can handle sign flips. A further feature is to turn the PLL off until a minimum signal magnitude is reached. This maintains the frequency while scanning e.g. over the nodes. The orange curve in Fig. \ref{fig:traces}(b) shows the result of our method: The anti nodes are equal in magnitude and the imaginary part stays very low. This robust method (iii) thus nicely maps the mode, even in the presence of frequency drifts, nodes, and sign changes.

The fourth mode-mapping approach is performed with the NWA \cite{davidovikj_NL_graphene_mode_visualization}. Here, a full frequency response is measured at every point of the line trace, and its fitted maximum and phase (App. \ref{app:demod}) are used to reconstruct the modal amplitude. Similar to our PLL (iii), method (iv) is also capable of mapping a drifting mode accurately [Fig. \ref{fig:traces}(b,c), green]. However, as Fig. \ref{fig:traces}(d) shows, the NWA method is about ten times slower compared to all other methods. Although it can be considered the gold standard, method (iv) is too slow to do e.g. full 2D mode maps efficiently. Our method (iii) is thus the preferred technique.

\begin{figure}[hbt]
\includegraphics[width=1.0\columnwidth]{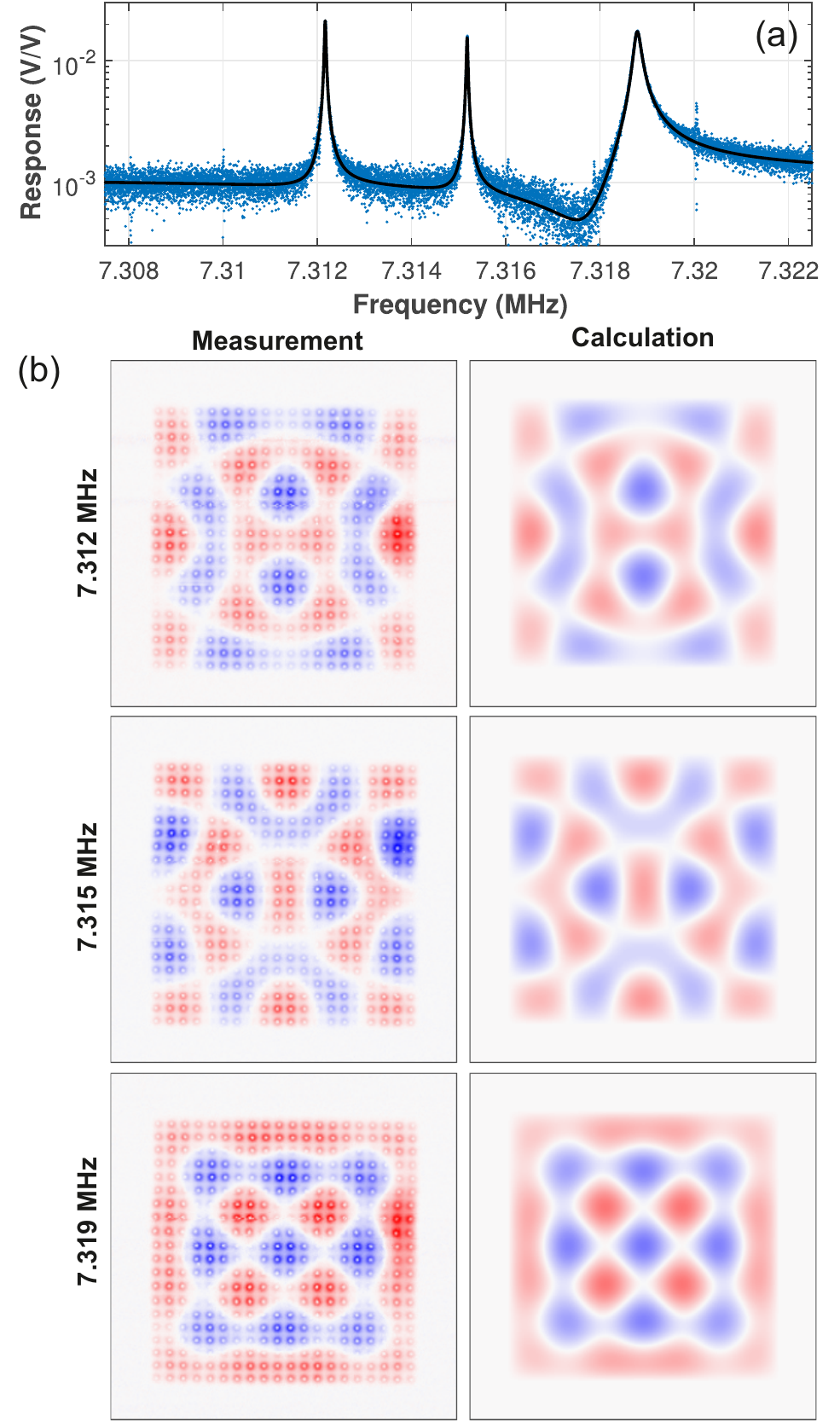}
\caption{(a) Driven response near $f_{5,5} = f_{1,7} = f_{7,1}$. The triplet is fitted using three standard harmonic oscillator responses taking crosstalk into account (App. \ref{app:demod}) \cite{poot_APL_cooling} (black line).
(b) Measured (left) and calculated (right) mode maps. The weights from the linear fitting are $\{w_{5,5},w_{1,7},w_{7,1}\} \propto \{0.561,   -0.506 ,0.656 \}$, $\{0.817,    0.260,   -0.515 \}$, and $\{ 0.105,    0.694,    0.712 \}$ for the left, middle, and right peak, respectively. The fit uncertainty in the weights is $ 0.002$. } \label{fig:triplet}
\end{figure}
To further demonstrate its use, full 2D maps of the first six modes were measured simultaneously as shown in Fig. \ref{fig:6modes}. The first mode at 1.46 MHz is indeed the (1,1) mode, and the second one at 2.30 MHz is the (1,2) mode. Although some modes are slightly distorted compared to Eq. \eqref{eq:u} (See App. \ref{app:modes} for discussion), one can still easily recognize them.
However, when modes are degenerate, a superposition of them is also an eigenmode and, a priory, it is not clear what their modes would look like. Mode mapping is thus crucial to understand the nature of the resonance. Eq. \eqref{eq:f} shows that the (5,5), (1,7), and (7,1) mode are triple degenerate and that these are expected around 7.3 MHz. Figure \ref{fig:triplet}(a) shows three distinct peaks. Their mode maps in Fig. \ref{fig:triplet}(b) indicate that, unlike the modes in Fig. \ref{fig:6modes}, their shapes are not directly given by Eq. \eqref{eq:u}; instead, they show a much richer spatial structure. With the resolution of our mapping, it is possible to quantitatively determine the contributions of each individual mode to the superpositions. For this, the mode shape is written as $u(x,y) = w_{5,5} u_{5,5}(x,y)+ w_{1,7}u_{1,7}(x,y)+w_{7,1}u_{7,1}(x,y)$ and the weights $w_{m,n}$ are determined by linear fitting to the experimental mode shapes. The results in the right column of Fig. \ref{fig:triplet}(b) show good agreement with the experiment, including the structure of nodal lines (white) and the variation in amplitude at the different anti nodes. Finally, note that the modes have very different damping rates $\gamma$, as seen from the peak width in Fig. \ref{fig:triplet}(a) (see also Table \ourtable). It is known that the clamping losses depend on the displacement field near the edge \cite{cole_natcomm_phonon_tunnelling, adiga_APL_SiN_drum_Q_mode, sun_NL_double_beam, singh_PRL_acoustic_blackbody_detection}. Looking at the first two mode shapes shows alternating positive (red) and negative (blue) displacements near the edge of the membrane, whereas the third one (cf. the one with increased damping) has the same sign everywhere along the edge (red only); the radiation of acoustical energy into the supports would be very different. This explanation for their different linewidths would be difficult to obtain without our high-resolution modes maps with phase information.

In conclusion, we have presented a robust method to efficiently map vibrational modes simultaneously and we have illustrated this technique using a high-stress \mySiN membrane. 

\begin{acknowledgments}
X. Yao assisted with the nanofabrication, and L. Rosendahl, E. Lebedev, and J. R\"owe with the setup and measurements. Funded by the German Research Foundation (DFG) under Germany's Excellence Strategy - EXC-2111-390814868 and TUM-IAS, funded by the German Excellence Initiative and the European Union Seventh Framework Programme under grant agreement 291763.
\end{acknowledgments}


\newpage
\appendix
\setcounter{section}{0}
\renewcommand\thesection{S\arabic{section}}
\renewcommand\thesubsection{S\arabic{section}.\arabic{subsection}}
\begin{huge}
Supplementary~Material
\end{huge}

\setcounter{figure}{0}
\renewcommand\thefigure{S\arabic{figure}}
\renewcommand{\thetable}{SI} 

\begin{table*}[htbp]
  \label{tab:modes}
  \centering
  \caption{Overview of the excitation power, setpoint, and fit parameters (see Fig. \ref{fig:zooms}) for the modes used in the main text. The fit uncertainty in the resonance frequency is below 3 Hz for all modes, but the exact value drifts over the course of time.}
  \begin{tabular}{ccccccc}
\textbf{Mode} & \textbf{Excitation } & \textbf{Frequency} & \boldmath{}\textbf{$\gamma / 2\pi $}\unboldmath{} & \boldmath{}\textbf{$z_\mathrm{max}$}\unboldmath{} & \boldmath{}\textbf{$\alpha$}\unboldmath{} & \boldmath{}\textbf{$\phi_\mathrm{sp}$}\unboldmath{} \\
      & \textbf{(dBm)} & \textbf{ (MHz)} & \textbf{(Hz)} & \textbf{(V/V)} & \textbf{(rad)} & \boldmath{}\textbf{$ \un{(\degree)}$}\unboldmath{} \\ \hline
(1,1) & -45   & 1.460 053 & $37.42 \pm 0.05$ & $0.527$ & $3.111 \pm 0.001$ & 87 \\
(1,2) & -50   & 2.304 098 & $9.22 \pm 0.09$ & $0.149$ & $0.594 \pm 0.007$ & -68 \\
(2,1) & -50   & 2.307 084 & $13.78 \pm 0.04$ & $0.384$ & $0.012 \pm 0.002$ & -98 \\
(2,2) & -35   & 2.915 386 & $14.00 \pm 0.26$ & $3.5 \cdot 10^{-3}$ & $2.283 \pm 0.014$ & -104 \\
(3,1) & -35   & 3.259 232 & $23.24 \pm 0.09$ & $0.030$ & $2.856 \pm 0.003$ & -111 \\
(1,3) & -40   & 3.260 257 & $31.14 \pm 0.05$ & $0.044$ & $2.788 \pm 0.001$ & -20 \\
Triplet L & -35   & 7.313 864 & $36.61 \pm 1.76$ & $9.1 \cdot 10^{-3}$ & $0.172 \pm 0.037$ & -68.8 \\
Triplet C & -35   & 7.316 988 & $30.74 \pm 1.58$ & $9.7 \cdot 10^{-3}$ & $2.842 \pm 0.039$ & -263.3 \\
Triplet R & -35   & 7.320 566 & $162.81 \pm 4.66$ & $0.011$ & $2.262 \pm 0.029$ & 49.8 \\
\end{tabular}%
\end{table*}%

\begin{figure}[htbp]
  \includegraphics[width=1.0\columnwidth]{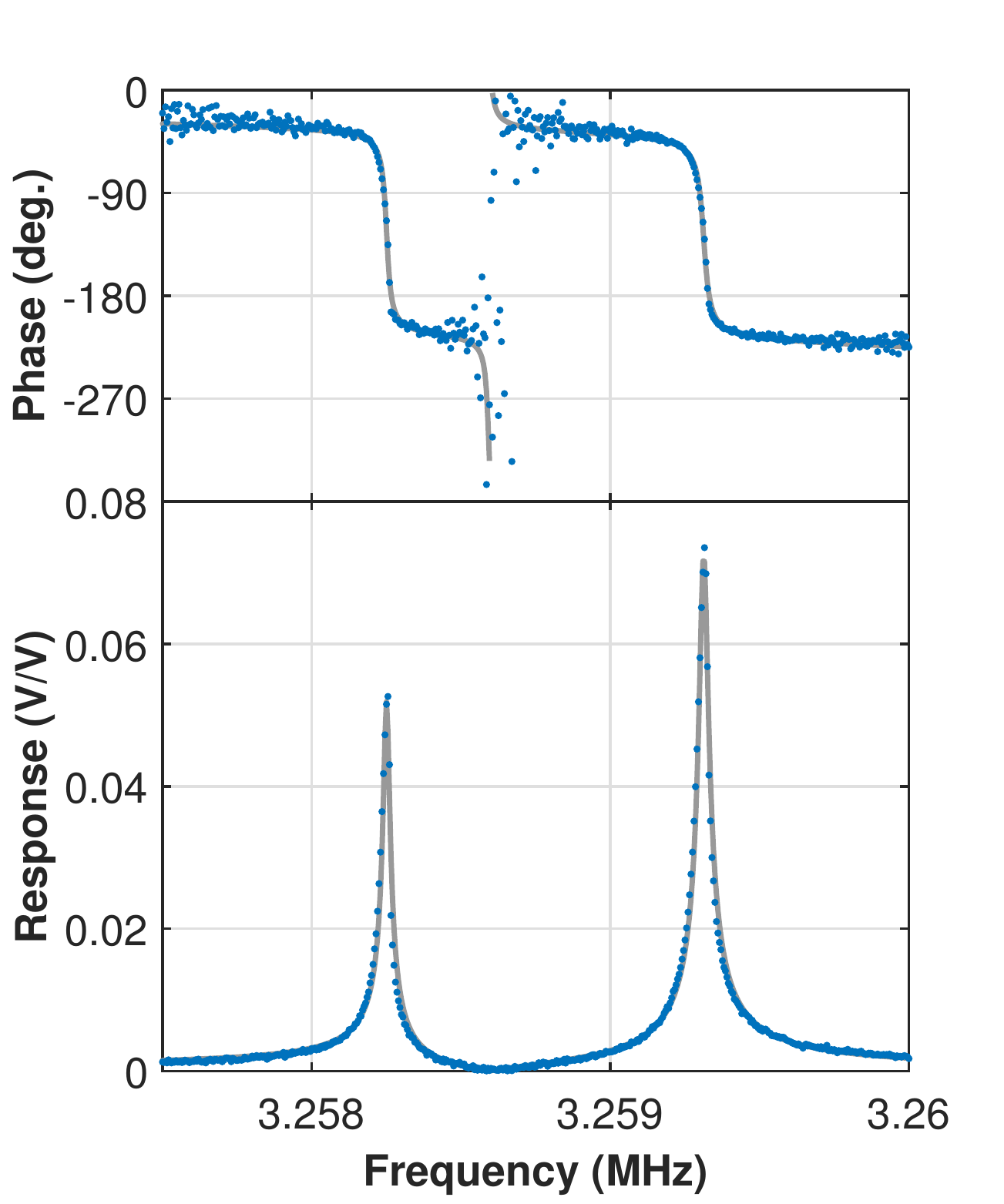}
  \caption{Network analyzer measurement of the driven response near the (3,1) and (1,3) eigenfrequencies. Two separate resonances with a spacing of 1.06 kHz can be seen. \label{fig:31_and_13}}%
\end{figure}

\begin{figure*}[htbp]
  \includegraphics[width=1.0\textwidth]{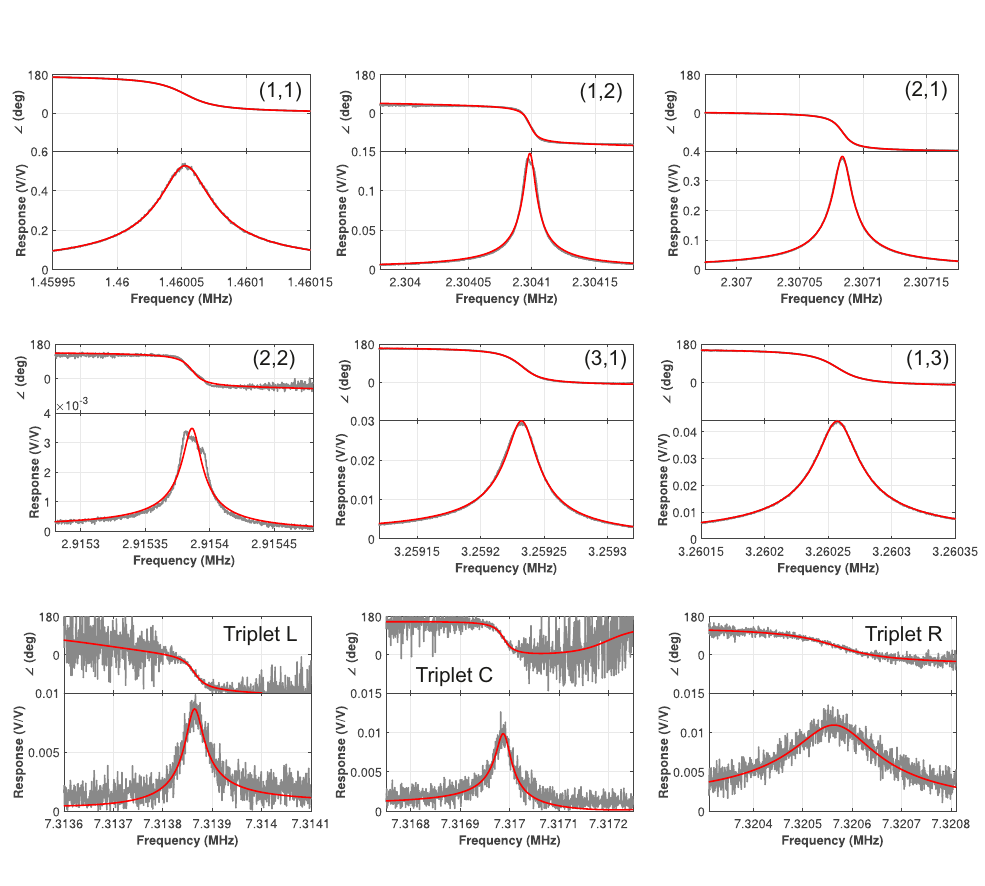}
  \caption{Driven responses of the modes. For the first 6 modes, the span is 200 Hz, whereas for the triplet the span is 500 Hz. The gray curves are measured using the LIA, whereas the red curves are the fits of Eq. (\ref{eq:H}) to the data. The measurement settings and fit results are indicated in Table \ourtable.}
  \label{fig:zooms}
\end{figure*}

\section{Demodulation, response, and mode shape} 
\label{app:demod}
In this section the details on the actuation and detection, lock-in operation and demodulation, response function, and their relation to the mode maps are discussed. 

\subsection{Excitation and detection}
As illustrated in Fig. \ref{fig:intro}(b), the LIA outputs a voltage at (angular) frequency $\omega$ with amplitude $V_\mathrm{out}$: $V_\mathrm{output}(t) = V_\mathrm{out}\cos(\omega t)$. This is applied to the piezo, resulting in an oscillating inertial force on the membrane, which induces vibrations. These result in a modulation of the reflected power that is detected using the photo detector and converted to a voltage by the low-noise amplifier. This voltage goes to the input of the LIA and contains the same frequency as the output: $V_\mathrm{input}(t) = V_\mathrm{in}\cos(\omega t+\phi)$. Here, $V_\mathrm{in}$ is the amplitude at the excitation frequency $\omega$ and $\phi$ is the phase difference between output and input signals. 

\subsection{Demodulation}
Demodulation with the LIA can be seen as a calculation of the quadratures $V_X$ and $V_Y$ using $V_Z(t) = \langle 2 V_\mathrm{input}(t) \exp(-i\omega t)\rangle \equiv V_X + iV_Y$. Here, the angled brackets indicate low pass filtering using the demodulation bandwidth. The complex demodulated voltage $V_Z(t)$ can also be expressed in its magnitude and phase $V_Z(t) = |V_Z(t)|\exp(i\angle V_Z(t))$, where $\angle$ denotes the argument of a complex number. For the aforementioned signal $V_\mathrm{input}(t) = V_\mathrm{in}\cos(\omega t+\phi)$, one obtains $V_X = V_\mathrm{in}\cos(\phi)$ and $V_Y = V_\mathrm{in}\sin(\phi)$ so that $|V_Z(t)|$ is $V_\mathrm{in}$ and $\angle V_Z(t) = \phi$. Due to the low-pass filtering, the demodulated signals (``quadratures'') are only slowly (compared to $\omega$) varying in time; their sampled values are indicated in the main text with the index $n$.

\subsection{Response and mode shape}
In the linear response regime used in our experiments, the measured signal $V_\mathrm{input}(t)$ is proportional to the drive amplitude $V_\mathrm{out}$. Hence, the responses in the main text are normalized by $V_\mathrm{out}$ and their units are, thus, $V/V$. In particular, we define $Z \equiv X +iY \equiv V_Z/V_\mathrm{out}$. This overall system response can be written as a product of the frequency responses of the individual components: $Z(\omega) = G(\omega)R(\omega)~\pderl{P}{U_{m,n}}~H_{m,n}(\omega) A(\omega)$. Here, $A$, $R$, and $G$ are the responses of the piezo transduction (``actuation''), photo detector (``responsivity''), and amplifier (``transimpedance gain''), respectively. Although on larger scales, such as in the overview in Fig. \ref{fig:intro}(c), signatures of the frequency-dependent piezo response can be seen\cite{hoch_Sbeam}, in a narrow span around the resonance frequency of interest, $\omega_0$, all these responses are approximately constant and the $\omega$ dependence is omitted. In contrast, $H_{m,n}(\omega)$ is the harmonic oscillator response function that transduces the piezo vibrations via the inertial force into the vibrational amplitude $U_{m,n}$ of the $(m,n)$ mode, which varies strongly near the eigenfrequency $\omega_0 = \omega_{m,n}$.\footnote{In principle, the response would be a sum over all modes $H_\mathrm{total}(\omega) = \sum_{m,n} H_{m,n}(\omega)$ but for simplicity it is assumed here that only a single mode $(m,n)$ gets excited. In Fig. \ref{fig:triplet}(a) and Fig. \ref{fig:31_and_13} the combined responses of respectively three and two modes are fitted to the data.} Finally, $\pderl{P}{U_{m,n}}$ is the change of the reflected laser power with displacement of the entire mode $U_{m,n}$ \cite{poot_physrep_quantum_regime}. Importantly, this quantity depends on the position of the laser spot: it is largest at anti-nodes and zero at a node. Using Eq. \eqref{eq:u} and applying the chain rule, it can be rewritten as $\pderl{P}{U_{m,n}} = \pderl{P}{u} \times \xi_{m,n}(x,y)$. Now $\pderl{P}{u}$ is independent of the position as it quantifies the change in reflected power $P$ for a given displacement $u$ \emph{at the readout position} $(x,y)$
\footnote{Further refinements in the readout model could include the finite spot size of the laser by writing $\pderl{P}{U_{m,n}}$ as a two dimensional convolution of the beam shape and the mode shape.}.
Combining all this, gives $Z(x,y, \omega) = Ce^{i\alpha} H_{m,n}(\omega) \xi_{m,n}(x,y)$ for some proportionality constant $C$ and complex angle $\alpha$. This shows that the mapped normalized demodulated signal at constant frequency $Z(x,y~|~\omega)$ is directly proportional to the mode shape $\xi_{m,n}(x,y)$. Measuring $Z$ as function of the readout position $(x,y)$ using any of the methods in the main text thus enables mapping the mode shape $\xi_{m,n}$.

\subsection{Harmonic oscillator response}
The harmonic oscillator response function $H(\omega)$ has the largest magnitude at the resonance frequency $\omega = \omega_0$. \cite{poot_physrep_quantum_regime} There, its phase is $\angle H(\omega_0) = -\pi/2$ so that on resonance, the total phase of $Z(\omega_0)$ becomes $\alpha -\pi/2 \mod \pi$ where the last term comes from the sign of $C\xi_{m,n}(x,y)$. This phase is typically used as the setpoint for the PLL, $\phi_\mathrm{sp}$ as it locks $\omega$ to $\omega_0$ thereby maximizing the magnitude of the signal. For this reason, the value of $\alpha$ is needed for every mode, which is determined by fitting the driven frequency response $Z(\omega~|~x,y)$ measured at a fixed position. In particular, they were fitted using the expression:
\begin{equation}
    Z(\omega) = z_\mathrm{max}e^{i\alpha}\frac{\omega_0\gamma}{\omega_0^2-\omega^2+i\omega\gamma} + z_Xe^{i\phi_X},
    \label{eq:H}
\end{equation}
where $z_\mathrm{max}$ is the peak value and $\gamma$ the damping rate which is related to the quality factor $Q = \omega_0/\gamma$. The $z_X$ term includes a small amount of electrical crosstalk\cite{poot_APL_cooling, hoch_Sbeam} (cf. the gray line in Fig. \ref{fig:intro}(c)) which result in Fano-like resonances.

Both during the demodulation (as a setting in the LIA), as well as in postprocessing, an additional ``analyzer'' phase $\psi$ can be set. In the former case, this corresponds to $\exp(-i\omega t) \rightarrow \exp(-i\omega t - \psi)$ for the aforementioned calculation of the quadratures. We use the latter, though, which performs the transformation
\begin{equation}
\left(\begin{array}{c}
X'\\Y'
\end{array}\right) = \left(\begin{array}{cc}
\cos \psi & \sin \psi \\ -\sin \psi & \cos \psi
\end{array}\right) \left(\begin{array}{c}
X\\Y
\end{array}\right).    
\end{equation}
By setting the analyzer phase equal to the setpoint of the phase-lock loop, $\psi = \phi_\mathrm{sp}$, the $X'$ quadrature (cf. the real part of $Z'$) contains the mode shape, whereas $Y'$ (cf. the imaginary part of $Z'$) is related to the error $e_j$.

So far the analysis has been for a single frequency $\omega$. The Zurich Instruments HF2 lock-in amplifier with multi-frequency kit can, however, generate an excitation signal with six different frequencies and subsequently demodulate the input signal at all of these. This enables measurements on six modes \emph{simultaneously}, as employed in, for example, Fig. \ref{fig:6modes}.

\section{Mode properties}
\label{app:modes}
In this Section, additional data on the different modes of the membrane are given. 

A zoom of the response near the (3,1) and (1,3) modes is shown in Fig. \ref{fig:31_and_13}. Figure \ref{fig:zooms} shows zooms of the modes discussed in the main text and Table \ourtable lists their properties. It is noted that due to temperature variations in the lab, the values for the exact resonance frequencies may differ slightly between measurements. This drift, in combination with the relatively long duration of response function measurements resulted in a clear distortion of the (2,2) mode. A similar measurement with the NWA showed a regular response and yielded $\gamma/2\pi = 5.22 \pm 0.02$, which corresponds to $Q = 559\mathrm{k}$.

Figure \ref{fig:6modes} showed the measured two dimensional mode maps of the first six modes of the square membrane. Some of the maps do not show completely straight nodal lines as expected from the theory. This may be an effect of the finite bending rigidity of the membrane \cite{jhung_NET_vibrations_perforated_plates}, and is confirmed using finite-element simulations. Another observation, is that $f_{1,2} < f_{2,1}$ whereas $f_{1,3} > f_{3,1}$. This means that the breaking of the degeneracy between the $(m,n)$ and $(n,m)$ modes for $m \neq n$ is not caused by different side lengths of the membrane (cf. a rectangular instead of a square shape) and that more subtle effects play a role here.
\bibliography{PLLmm}

\begin{thebibliography}{36}%
\makeatletter
\providecommand \@ifxundefined [1]{%
 \@ifx{#1\undefined}
}%
\providecommand \@ifnum [1]{%
 \ifnum #1\expandafter \@firstoftwo
 \else \expandafter \@secondoftwo
 \fi
}%
\providecommand \@ifx [1]{%
 \ifx #1\expandafter \@firstoftwo
 \else \expandafter \@secondoftwo
 \fi
}%
\providecommand \natexlab [1]{#1}%
\providecommand \enquote  [1]{``#1''}%
\providecommand \bibnamefont  [1]{#1}%
\providecommand \bibfnamefont [1]{#1}%
\providecommand \citenamefont [1]{#1}%
\providecommand \href@noop [0]{\@secondoftwo}%
\providecommand \href [0]{\begingroup \@sanitize@url \@href}%
\providecommand \@href[1]{\@@startlink{#1}\@@href}%
\providecommand \@@href[1]{\endgroup#1\@@endlink}%
\providecommand \@sanitize@url [0]{\catcode `\\12\catcode `\$12\catcode
  `\&12\catcode `\#12\catcode `\^12\catcode `\_12\catcode `\%12\relax}%
\providecommand \@@startlink[1]{}%
\providecommand \@@endlink[0]{}%
\providecommand \url  [0]{\begingroup\@sanitize@url \@url }%
\providecommand \@url [1]{\endgroup\@href {#1}{\urlprefix }}%
\providecommand \urlprefix  [0]{URL }%
\providecommand \Eprint [0]{\href }%
\providecommand \doibase [0]{http://dx.doi.org/}%
\providecommand \selectlanguage [0]{\@gobble}%
\providecommand \bibinfo  [0]{\@secondoftwo}%
\providecommand \bibfield  [0]{\@secondoftwo}%
\providecommand \translation [1]{[#1]}%
\providecommand \BibitemOpen [0]{}%
\providecommand \bibitemStop [0]{}%
\providecommand \bibitemNoStop [0]{.\EOS\space}%
\providecommand \EOS [0]{\spacefactor3000\relax}%
\providecommand \BibitemShut  [1]{\csname bibitem#1\endcsname}%
\let\auto@bib@innerbib\@empty
\bibitem [{\citenamefont {Hesjedal}, \citenamefont {Chilla},\ and\
  \citenamefont {Fröhlich}(1997)}]{hesjedal_APL_AFM_SAW}%
  \BibitemOpen
  \bibfield  {author} {\bibinfo {author} {\bibfnamefont {T.}~\bibnamefont
  {Hesjedal}}, \bibinfo {author} {\bibfnamefont {E.}~\bibnamefont {Chilla}}, \
  and\ \bibinfo {author} {\bibfnamefont {H.-J.}\ \bibnamefont {Fröhlich}},\
  }\bibfield  {title} {\enquote {\bibinfo {title} {High resolution
  visualization of acoustic wave fields within surface acoustic wave
  devices},}\ }\href {\doibase 10.1063/1.119323} {\bibfield  {journal}
  {\bibinfo  {journal} {Appl. Phys. Lett.}\ }\textbf {\bibinfo {volume} {70}},\
  \bibinfo {pages} {1372--1374} (\bibinfo {year} {1997})},\ \Eprint
  {http://arxiv.org/abs/https://doi.org/10.1063/1.119323}
  {https://doi.org/10.1063/1.119323} \BibitemShut {NoStop}%
\bibitem [{\citenamefont {Venkatasubramanian}\ \emph
  {et~al.}(2016)\citenamefont {Venkatasubramanian}, \citenamefont {Sauer},
  \citenamefont {Roy}, \citenamefont {Xia}, \citenamefont {Wishart},\ and\
  \citenamefont {Hiebert}}]{masssensor}%
  \BibitemOpen
  \bibfield  {author} {\bibinfo {author} {\bibfnamefont {A.}~\bibnamefont
  {Venkatasubramanian}}, \bibinfo {author} {\bibfnamefont {V.~T.~K.}\
  \bibnamefont {Sauer}}, \bibinfo {author} {\bibfnamefont {S.~K.}\ \bibnamefont
  {Roy}}, \bibinfo {author} {\bibfnamefont {M.}~\bibnamefont {Xia}}, \bibinfo
  {author} {\bibfnamefont {D.~S.}\ \bibnamefont {Wishart}}, \ and\ \bibinfo
  {author} {\bibfnamefont {W.~K.}\ \bibnamefont {Hiebert}},\ }\bibfield
  {title} {\enquote {\bibinfo {title} {Nano-optomechanical systems for gas
  chromatography},}\ }\href {\doibase 10.1021/acs.nanolett.6b03066} {\bibfield
  {journal} {\bibinfo  {journal} {Nano Letters}\ }\textbf {\bibinfo {volume}
  {16}},\ \bibinfo {pages} {6975--6981} (\bibinfo {year} {2016})}\BibitemShut
  {NoStop}%
\bibitem [{\citenamefont {Naik}\ \emph {et~al.}(2009)\citenamefont {Naik},
  \citenamefont {Hanay}, \citenamefont {Hiebert}, \citenamefont {Feng},\ and\
  \citenamefont {Roukes}}]{naik_natnano_masssensing}%
  \BibitemOpen
  \bibfield  {author} {\bibinfo {author} {\bibfnamefont {A.~K.}\ \bibnamefont
  {Naik}}, \bibinfo {author} {\bibfnamefont {M.~S.}\ \bibnamefont {Hanay}},
  \bibinfo {author} {\bibfnamefont {W.~K.}\ \bibnamefont {Hiebert}}, \bibinfo
  {author} {\bibfnamefont {X.~L.}\ \bibnamefont {Feng}}, \ and\ \bibinfo
  {author} {\bibfnamefont {M.~L.}\ \bibnamefont {Roukes}},\ }\bibfield  {title}
  {\enquote {\bibinfo {title} {Towards single-molecule nanomechanical mass
  spectrometry},}\ }\href {http://dx.doi.org/10.1038/nnano.2009.152} {\bibfield
   {journal} {\bibinfo  {journal} {Nat Nano}\ }\textbf {\bibinfo {volume}
  {4}},\ \bibinfo {pages} {445--450} (\bibinfo {year} {2009})}\BibitemShut
  {NoStop}%
\bibitem [{\citenamefont {McKeown}\ \emph {et~al.}(2017)\citenamefont
  {McKeown}, \citenamefont {Wang}, \citenamefont {Yu},\ and\ \citenamefont
  {Goddard}}]{mckeown_MNE_gassensor}%
  \BibitemOpen
  \bibfield  {author} {\bibinfo {author} {\bibfnamefont {S.~J.}\ \bibnamefont
  {McKeown}}, \bibinfo {author} {\bibfnamefont {X.}~\bibnamefont {Wang}},
  \bibinfo {author} {\bibfnamefont {X.}~\bibnamefont {Yu}}, \ and\ \bibinfo
  {author} {\bibfnamefont {L.~L.}\ \bibnamefont {Goddard}},\ }\bibfield
  {title} {\enquote {\bibinfo {title} {Realization of palladium-based
  optomechanical cantilever hydrogen sensor},}\ }\href {\doibase
  https://doi.org/10.1038/micronano.2016.87} {\bibfield  {journal} {\bibinfo
  {journal} {Microsyst Nanoneng}\ }\textbf {\bibinfo {volume} {3}} (\bibinfo
  {year} {2017}),\ https://doi.org/10.1038/micronano.2016.87}\BibitemShut
  {NoStop}%
\bibitem [{\citenamefont {Melcher}\ \emph {et~al.}(2014)\citenamefont
  {Melcher}, \citenamefont {Stirling}, \citenamefont {Cervantes}, \citenamefont
  {Pratt},\ and\ \citenamefont {Shaw}}]{forcesensor}%
  \BibitemOpen
  \bibfield  {author} {\bibinfo {author} {\bibfnamefont {J.}~\bibnamefont
  {Melcher}}, \bibinfo {author} {\bibfnamefont {J.}~\bibnamefont {Stirling}},
  \bibinfo {author} {\bibfnamefont {F.~G.}\ \bibnamefont {Cervantes}}, \bibinfo
  {author} {\bibfnamefont {J.~R.}\ \bibnamefont {Pratt}}, \ and\ \bibinfo
  {author} {\bibfnamefont {G.~A.}\ \bibnamefont {Shaw}},\ }\bibfield  {title}
  {\enquote {\bibinfo {title} {A self-calibrating optomechanical force sensor
  with femtonewton resolution},}\ }\href {\doibase
  https://doi.org/10.1063/1.4903801} {\bibfield  {journal} {\bibinfo  {journal}
  {Appl. Phys. Lett.}\ }\textbf {\bibinfo {volume} {105}} (\bibinfo {year}
  {2014}),\ https://doi.org/10.1063/1.4903801}\BibitemShut {NoStop}%
\bibitem [{\citenamefont {Bleszynski-Jayich}\ \emph {et~al.}(2009)\citenamefont
  {Bleszynski-Jayich}, \citenamefont {Shanks}, \citenamefont {Peaudecerf},
  \citenamefont {Ginossar}, \citenamefont {von Oppen}, \citenamefont
  {Glazman},\ and\ \citenamefont
  {Harris}}]{bleszynski-jayich_science_persistent_currents}%
  \BibitemOpen
  \bibfield  {author} {\bibinfo {author} {\bibfnamefont {A.~C.}\ \bibnamefont
  {Bleszynski-Jayich}}, \bibinfo {author} {\bibfnamefont {W.~E.}\ \bibnamefont
  {Shanks}}, \bibinfo {author} {\bibfnamefont {B.}~\bibnamefont {Peaudecerf}},
  \bibinfo {author} {\bibfnamefont {E.}~\bibnamefont {Ginossar}}, \bibinfo
  {author} {\bibfnamefont {F.}~\bibnamefont {von Oppen}}, \bibinfo {author}
  {\bibfnamefont {L.}~\bibnamefont {Glazman}}, \ and\ \bibinfo {author}
  {\bibfnamefont {J.~G.~E.}\ \bibnamefont {Harris}},\ }\bibfield  {title}
  {\enquote {\bibinfo {title} {Persistent currents in normal metal rings},}\
  }\href {http://www.sciencemag.org/cgi/content/abstract/326/5950/272}
  {\bibfield  {journal} {\bibinfo  {journal} {Science}\ }\textbf {\bibinfo
  {volume} {326}},\ \bibinfo {pages} {272--275} (\bibinfo {year}
  {2009})}\BibitemShut {NoStop}%
\bibitem [{\citenamefont {Fong}\ \emph {et~al.}(2019)\citenamefont {Fong},
  \citenamefont {Li}, \citenamefont {Zhao}, \citenamefont {Yang}, \citenamefont
  {Wang},\ and\ \citenamefont {Zhang}}]{fong_nature_casimir}%
  \BibitemOpen
  \bibfield  {author} {\bibinfo {author} {\bibfnamefont {K.~Y.}\ \bibnamefont
  {Fong}}, \bibinfo {author} {\bibfnamefont {H.-K.}\ \bibnamefont {Li}},
  \bibinfo {author} {\bibfnamefont {R.}~\bibnamefont {Zhao}}, \bibinfo {author}
  {\bibfnamefont {S.}~\bibnamefont {Yang}}, \bibinfo {author} {\bibfnamefont
  {Y.}~\bibnamefont {Wang}}, \ and\ \bibinfo {author} {\bibfnamefont
  {X.}~\bibnamefont {Zhang}},\ }\bibfield  {title} {\enquote {\bibinfo {title}
  {Phonon heat transfer across a vacuum through quantum fluctuations},}\ }\href
  {\doibase 10.1038/s41586-019-1800-4} {\bibfield  {journal} {\bibinfo
  {journal} {Nature}\ }\textbf {\bibinfo {volume} {576}},\ \bibinfo {pages}
  {243--247} (\bibinfo {year} {2019})}\BibitemShut {NoStop}%
\bibitem [{\citenamefont {Singh}\ and\ \citenamefont
  {Purdy}(2020)}]{singh_PRL_acoustic_blackbody_detection}%
  \BibitemOpen
  \bibfield  {author} {\bibinfo {author} {\bibfnamefont {R.}~\bibnamefont
  {Singh}}\ and\ \bibinfo {author} {\bibfnamefont {T.~P.}\ \bibnamefont
  {Purdy}},\ }\bibfield  {title} {\enquote {\bibinfo {title} {Detecting
  acoustic blackbody radiation with an optomechanical antenna},}\ }\href
  {\doibase 10.1103/PhysRevLett.125.120603} {\bibfield  {journal} {\bibinfo
  {journal} {Phys. Rev. Lett.}\ }\textbf {\bibinfo {volume} {125}},\ \bibinfo
  {pages} {120603} (\bibinfo {year} {2020})}\BibitemShut {NoStop}%
\bibitem [{\citenamefont {LaHaye}\ \emph {et~al.}(2004)\citenamefont {LaHaye},
  \citenamefont {Buu}, \citenamefont {Camarota},\ and\ \citenamefont
  {Schwab}}]{lahaye_science_quantum_limit}%
  \BibitemOpen
  \bibfield  {author} {\bibinfo {author} {\bibfnamefont {M.~D.}\ \bibnamefont
  {LaHaye}}, \bibinfo {author} {\bibfnamefont {O.}~\bibnamefont {Buu}},
  \bibinfo {author} {\bibfnamefont {B.}~\bibnamefont {Camarota}}, \ and\
  \bibinfo {author} {\bibfnamefont {K.~C.}\ \bibnamefont {Schwab}},\ }\bibfield
   {title} {\enquote {\bibinfo {title} {Approaching the quantum limit of a
  nanomechanical resonator},}\ }\href
  {http://www.sciencemag.org/cgi/content/abstract/304/5667/74} {\bibfield
  {journal} {\bibinfo  {journal} {Science}\ }\textbf {\bibinfo {volume}
  {304}},\ \bibinfo {pages} {74--77} (\bibinfo {year} {2004})}\BibitemShut
  {NoStop}%
\bibitem [{\citenamefont {Etaki}\ \emph {et~al.}(2008)\citenamefont {Etaki},
  \citenamefont {Poot}, \citenamefont {Mahboob}, \citenamefont {Onomitsu},
  \citenamefont {Yamaguchi},\ and\ \citenamefont {van~der
  Zant}}]{etaki_natphys_squid}%
  \BibitemOpen
  \bibfield  {author} {\bibinfo {author} {\bibfnamefont {S.}~\bibnamefont
  {Etaki}}, \bibinfo {author} {\bibfnamefont {M.}~\bibnamefont {Poot}},
  \bibinfo {author} {\bibfnamefont {I.}~\bibnamefont {Mahboob}}, \bibinfo
  {author} {\bibfnamefont {K.}~\bibnamefont {Onomitsu}}, \bibinfo {author}
  {\bibfnamefont {H.}~\bibnamefont {Yamaguchi}}, \ and\ \bibinfo {author}
  {\bibfnamefont {H.~S.~J.}\ \bibnamefont {van~der Zant}},\ }\bibfield  {title}
  {\enquote {\bibinfo {title} {Motion detection of a micromechanical resonator
  embedded in a d.c. squid},}\ }\href {http://dx.doi.org/10.1038/nphys1057}
  {\bibfield  {journal} {\bibinfo  {journal} {Nature Physics}\ }\textbf
  {\bibinfo {volume} {4}},\ \bibinfo {pages} {785--788} (\bibinfo {year}
  {2008})}\BibitemShut {NoStop}%
\bibitem [{\citenamefont {Anetsberger}\ \emph {et~al.}(2009)\citenamefont
  {Anetsberger}, \citenamefont {Arcizet}, \citenamefont {Unterreithmeier},
  \citenamefont {Riviere}, \citenamefont {Schliesser}, \citenamefont {Weig},
  \citenamefont {Kotthaus},\ and\ \citenamefont
  {Kippenberg}}]{anetsberger_natphys_nearfield}%
  \BibitemOpen
  \bibfield  {author} {\bibinfo {author} {\bibfnamefont {G.}~\bibnamefont
  {Anetsberger}}, \bibinfo {author} {\bibfnamefont {O.}~\bibnamefont
  {Arcizet}}, \bibinfo {author} {\bibfnamefont {Q.~P.}\ \bibnamefont
  {Unterreithmeier}}, \bibinfo {author} {\bibfnamefont {R.}~\bibnamefont
  {Riviere}}, \bibinfo {author} {\bibfnamefont {A.}~\bibnamefont {Schliesser}},
  \bibinfo {author} {\bibfnamefont {E.~M.}\ \bibnamefont {Weig}}, \bibinfo
  {author} {\bibfnamefont {J.~P.}\ \bibnamefont {Kotthaus}}, \ and\ \bibinfo
  {author} {\bibfnamefont {T.~J.}\ \bibnamefont {Kippenberg}},\ }\bibfield
  {title} {\enquote {\bibinfo {title} {Near-field cavity optomechanics with
  nanomechanical oscillators},}\ }\href {http://dx.doi.org/10.1038/nphys1425}
  {\bibfield  {journal} {\bibinfo  {journal} {Nat Phys}\ }\textbf {\bibinfo
  {volume} {5}},\ \bibinfo {pages} {909--914} (\bibinfo {year}
  {2009})}\BibitemShut {NoStop}%
\bibitem [{\citenamefont {Barg}\ \emph {et~al.}(2016)\citenamefont {Barg},
  \citenamefont {Tsaturyan}, \citenamefont {Belhage}, \citenamefont {Nielsen},
  \citenamefont {Møller},\ and\ \citenamefont
  {Schliesser}}]{barg_APB_darkfield_mode_imaging}%
  \BibitemOpen
  \bibfield  {author} {\bibinfo {author} {\bibfnamefont {A.}~\bibnamefont
  {Barg}}, \bibinfo {author} {\bibfnamefont {Y.}~\bibnamefont {Tsaturyan}},
  \bibinfo {author} {\bibfnamefont {E.}~\bibnamefont {Belhage}}, \bibinfo
  {author} {\bibfnamefont {W.~H.~P.}\ \bibnamefont {Nielsen}}, \bibinfo
  {author} {\bibfnamefont {C.~B.}\ \bibnamefont {Møller}}, \ and\ \bibinfo
  {author} {\bibfnamefont {A.}~\bibnamefont {Schliesser}},\ }\bibfield  {title}
  {\enquote {\bibinfo {title} {Measuring and imaging nanomechanical motion with
  laser light},}\ }\href {\doibase 10.1007/s00340-016-6585-7} {\bibfield
  {journal} {\bibinfo  {journal} {Applied Physics B}\ }\textbf {\bibinfo
  {volume} {123}},\ \bibinfo {pages} {8} (\bibinfo {year} {2016})}\BibitemShut
  {NoStop}%
\bibitem [{\citenamefont {Zhang}\ \emph {et~al.}(2015)\citenamefont {Zhang},
  \citenamefont {Waitz}, \citenamefont {Yang}, \citenamefont {Lutz},
  \citenamefont {Angelova}, \citenamefont {Gölzhäuser},\ and\ \citenamefont
  {Scheer}}]{zhang_APL_ultrathin_membranes}%
  \BibitemOpen
  \bibfield  {author} {\bibinfo {author} {\bibfnamefont {X.}~\bibnamefont
  {Zhang}}, \bibinfo {author} {\bibfnamefont {R.}~\bibnamefont {Waitz}},
  \bibinfo {author} {\bibfnamefont {F.}~\bibnamefont {Yang}}, \bibinfo {author}
  {\bibfnamefont {C.}~\bibnamefont {Lutz}}, \bibinfo {author} {\bibfnamefont
  {P.}~\bibnamefont {Angelova}}, \bibinfo {author} {\bibfnamefont
  {A.}~\bibnamefont {Gölzhäuser}}, \ and\ \bibinfo {author} {\bibfnamefont
  {E.}~\bibnamefont {Scheer}},\ }\bibfield  {title} {\enquote {\bibinfo {title}
  {Vibrational modes of ultrathin carbon nanomembrane mechanical resonators},}\
  }\href {https://pub.uni-bielefeld.de/record/2728386} {\bibfield  {journal}
  {\bibinfo  {journal} {Appl Phys Lett}\ }\textbf {\bibinfo {volume} {106}}
  (\bibinfo {year} {2015})}\BibitemShut {NoStop}%
\bibitem [{\citenamefont {Davidovikj}\ \emph {et~al.}(2016)\citenamefont
  {Davidovikj}, \citenamefont {Slim}, \citenamefont {Cartamil-Bueno},
  \citenamefont {van~der Zant}, \citenamefont {Steeneken},\ and\ \citenamefont
  {Venstra}}]{davidovikj_NL_graphene_mode_visualization}%
  \BibitemOpen
  \bibfield  {author} {\bibinfo {author} {\bibfnamefont {D.}~\bibnamefont
  {Davidovikj}}, \bibinfo {author} {\bibfnamefont {J.~J.}\ \bibnamefont
  {Slim}}, \bibinfo {author} {\bibfnamefont {S.~J.}\ \bibnamefont
  {Cartamil-Bueno}}, \bibinfo {author} {\bibfnamefont {H.~S.~J.}\ \bibnamefont
  {van~der Zant}}, \bibinfo {author} {\bibfnamefont {P.~G.}\ \bibnamefont
  {Steeneken}}, \ and\ \bibinfo {author} {\bibfnamefont {W.~J.}\ \bibnamefont
  {Venstra}},\ }\bibfield  {title} {\enquote {\bibinfo {title} {Visualizing the
  motion of graphene nanodrums},}\ }\href {\doibase
  10.1021/acs.nanolett.6b00477} {\bibfield  {journal} {\bibinfo  {journal}
  {Nano Lett.}\ }\textbf {\bibinfo {volume} {16}},\ \bibinfo {pages}
  {2768--2773} (\bibinfo {year} {2016})}\BibitemShut {NoStop}%
\bibitem [{\citenamefont {Shen}\ \emph {et~al.}(2017)\citenamefont {Shen},
  \citenamefont {Han}, \citenamefont {Zou},\ and\ \citenamefont
  {Tang}}]{shen_RSI_image_AlN_microdisk_vibrations}%
  \BibitemOpen
  \bibfield  {author} {\bibinfo {author} {\bibfnamefont {Z.}~\bibnamefont
  {Shen}}, \bibinfo {author} {\bibfnamefont {X.}~\bibnamefont {Han}}, \bibinfo
  {author} {\bibfnamefont {C.-L.}\ \bibnamefont {Zou}}, \ and\ \bibinfo
  {author} {\bibfnamefont {H.~X.}\ \bibnamefont {Tang}},\ }\bibfield  {title}
  {\enquote {\bibinfo {title} {Phase sensitive imaging of 10 ghz vibrations in
  an aln microdisk resonator},}\ }\href {\doibase 10.1063/1.4995008} {\bibfield
   {journal} {\bibinfo  {journal} {Review of scientific instruments}\ }\textbf
  {\bibinfo {volume} {88}} (\bibinfo {year} {2017}),\
  10.1063/1.4995008}\BibitemShut {NoStop}%
\bibitem [{\citenamefont {Romero}\ \emph {et~al.}(2019)\citenamefont {Romero},
  \citenamefont {Kalra}, \citenamefont {Mauranyapin}, \citenamefont {Baker},
  \citenamefont {Meng},\ and\ \citenamefont
  {Bowen}}]{Romero_PRAppl_SiN_phonon_waveguide_heterodyn}%
  \BibitemOpen
  \bibfield  {author} {\bibinfo {author} {\bibfnamefont {E.}~\bibnamefont
  {Romero}}, \bibinfo {author} {\bibfnamefont {R.}~\bibnamefont {Kalra}},
  \bibinfo {author} {\bibfnamefont {N.}~\bibnamefont {Mauranyapin}}, \bibinfo
  {author} {\bibfnamefont {C.}~\bibnamefont {Baker}}, \bibinfo {author}
  {\bibfnamefont {C.}~\bibnamefont {Meng}}, \ and\ \bibinfo {author}
  {\bibfnamefont {W.}~\bibnamefont {Bowen}},\ }\bibfield  {title} {\enquote
  {\bibinfo {title} {Propagation and imaging of mechanical waves in a highly
  stressed single-mode acoustic waveguide},}\ }\href {\doibase
  10.1103/PhysRevApplied.11.064035} {\bibfield  {journal} {\bibinfo  {journal}
  {Phys. Rev. Applied}\ }\textbf {\bibinfo {volume} {11}},\ \bibinfo {pages}
  {064035} (\bibinfo {year} {2019})}\BibitemShut {NoStop}%
\bibitem [{\citenamefont {Garcia-Sanchez}\ \emph {et~al.}(2008)\citenamefont
  {Garcia-Sanchez}, \citenamefont {van~der Zande}, \citenamefont {Paulo},
  \citenamefont {Lassagne}, \citenamefont {McEuen},\ and\ \citenamefont
  {Bachtold}}]{garcia_NL_imaging_graphene}%
  \BibitemOpen
  \bibfield  {author} {\bibinfo {author} {\bibfnamefont {D.}~\bibnamefont
  {Garcia-Sanchez}}, \bibinfo {author} {\bibfnamefont {A.~M.}\ \bibnamefont
  {van~der Zande}}, \bibinfo {author} {\bibfnamefont {A.~S.}\ \bibnamefont
  {Paulo}}, \bibinfo {author} {\bibfnamefont {B.}~\bibnamefont {Lassagne}},
  \bibinfo {author} {\bibfnamefont {P.~L.}\ \bibnamefont {McEuen}}, \ and\
  \bibinfo {author} {\bibfnamefont {A.}~\bibnamefont {Bachtold}},\ }\bibfield
  {title} {\enquote {\bibinfo {title} {Imaging mechanical vibrations in
  suspended graphene sheets},}\ }\href {http://dx.doi.org/10.1021/nl080201h}
  {\bibfield  {journal} {\bibinfo  {journal} {Nano Letters}\ }\textbf {\bibinfo
  {volume} {8}},\ \bibinfo {pages} {1399--1403} (\bibinfo {year}
  {2008})}\BibitemShut {NoStop}%
\bibitem [{\citenamefont {Garcia-Sanchez}\ \emph {et~al.}(2007)\citenamefont
  {Garcia-Sanchez}, \citenamefont {Paulo}, \citenamefont {Esplandiu},
  \citenamefont {Perez-Murano}, \citenamefont {Forr\'{o}}, \citenamefont
  {Aguasca},\ and\ \citenamefont {Bachtold}}]{garcia_PRL_dfm_nanotube}%
  \BibitemOpen
  \bibfield  {author} {\bibinfo {author} {\bibfnamefont {D.}~\bibnamefont
  {Garcia-Sanchez}}, \bibinfo {author} {\bibfnamefont {A.~S.}\ \bibnamefont
  {Paulo}}, \bibinfo {author} {\bibfnamefont {M.~J.}\ \bibnamefont
  {Esplandiu}}, \bibinfo {author} {\bibfnamefont {F.}~\bibnamefont
  {Perez-Murano}}, \bibinfo {author} {\bibfnamefont {L.}~\bibnamefont
  {Forr\'{o}}}, \bibinfo {author} {\bibfnamefont {A.}~\bibnamefont {Aguasca}},
  \ and\ \bibinfo {author} {\bibfnamefont {A.}~\bibnamefont {Bachtold}},\
  }\bibfield  {title} {\enquote {\bibinfo {title} {Mechanical detection of
  carbon nanotube resonator vibrations},}\ }\href {\doibase
  10.1103/PhysRevLett.99.085501} {\bibfield  {journal} {\bibinfo  {journal}
  {Phys. Rev. Lett.}\ }\textbf {\bibinfo {volume} {99}},\ \bibinfo {eid}
  {085501} (\bibinfo {year} {2007})}\BibitemShut {NoStop}%
\bibitem [{\citenamefont {Hoch}\ \emph {et~al.}(2020)\citenamefont {Hoch},
  \citenamefont {Sommer}, \citenamefont {Mueller},\ and\ \citenamefont
  {Poot}}]{hoch_TJP_onchip}%
  \BibitemOpen
  \bibfield  {author} {\bibinfo {author} {\bibfnamefont {D.}~\bibnamefont
  {Hoch}}, \bibinfo {author} {\bibfnamefont {T.}~\bibnamefont {Sommer}},
  \bibinfo {author} {\bibfnamefont {S.}~\bibnamefont {Mueller}}, \ and\
  \bibinfo {author} {\bibfnamefont {M.}~\bibnamefont {Poot}},\ }\bibfield
  {title} {\enquote {\bibinfo {title} {On-chip quantum optics and integrated
  optomechanics},}\ }\href {\doibase 10.3906/fiz-2004-20} {\bibfield  {journal}
  {\bibinfo  {journal} {Turkish Journal of Physics}\ }\textbf {\bibinfo
  {volume} {44}},\ \bibinfo {pages} {239 -- 246} (\bibinfo {year}
  {2020})}\BibitemShut {NoStop}%
\bibitem [{\citenamefont {Hoch}, \citenamefont {Yao},\ and\ \citenamefont
  {Poot}(2020)}]{hoch_Sbeam}%
  \BibitemOpen
  \bibfield  {author} {\bibinfo {author} {\bibfnamefont {D.}~\bibnamefont
  {Hoch}}, \bibinfo {author} {\bibfnamefont {X.}~\bibnamefont {Yao}}, \ and\
  \bibinfo {author} {\bibfnamefont {M.}~\bibnamefont {Poot}},\ }\href@noop {}
  {\enquote {\bibinfo {title} {Geometric tuning of stress in silicon nitride
  beam resonators},}\ } (\bibinfo {year} {2020}),\ \bibinfo {note} {in
  preparation}\BibitemShut {NoStop}%
\bibitem [{\citenamefont {Terrasanta}\ \emph {et~al.}(2021)\citenamefont
  {Terrasanta}, \citenamefont {M{\"u}ller}, \citenamefont {Sommer},
  \citenamefont {Gepr\"{a}gs}, \citenamefont {Gross}, \citenamefont
  {Althammer},\ and\ \citenamefont {Poot}}]{terrasanta_AlN_on_SiN}%
  \BibitemOpen
  \bibfield  {author} {\bibinfo {author} {\bibfnamefont {G.}~\bibnamefont
  {Terrasanta}}, \bibinfo {author} {\bibfnamefont {M.}~\bibnamefont
  {M{\"u}ller}}, \bibinfo {author} {\bibfnamefont {T.}~\bibnamefont {Sommer}},
  \bibinfo {author} {\bibfnamefont {S.}~\bibnamefont {Gepr\"{a}gs}}, \bibinfo
  {author} {\bibfnamefont {R.}~\bibnamefont {Gross}}, \bibinfo {author}
  {\bibfnamefont {M.}~\bibnamefont {Althammer}}, \ and\ \bibinfo {author}
  {\bibfnamefont {M.}~\bibnamefont {Poot}},\ }\href@noop {} {\enquote {\bibinfo
  {title} {{Growth of Aluminum Nitride on a Silicon Nitride Substrate for
  Hybrid Photonic Circuits}},}\ } (\bibinfo {year} {2021}),\ \bibinfo {note}
  {arXiv:2103.08318}\BibitemShut {NoStop}%
\bibitem [{\citenamefont {Adiga}\ \emph {et~al.}(2011)\citenamefont {Adiga},
  \citenamefont {Ilic}, \citenamefont {Barton}, \citenamefont {Wilson-Rae},
  \citenamefont {Craighead},\ and\ \citenamefont
  {Parpia}}]{adiga_APL_SiN_drum_Q_mode}%
  \BibitemOpen
  \bibfield  {author} {\bibinfo {author} {\bibfnamefont {V.~P.}\ \bibnamefont
  {Adiga}}, \bibinfo {author} {\bibfnamefont {B.}~\bibnamefont {Ilic}},
  \bibinfo {author} {\bibfnamefont {R.~A.}\ \bibnamefont {Barton}}, \bibinfo
  {author} {\bibfnamefont {I.}~\bibnamefont {Wilson-Rae}}, \bibinfo {author}
  {\bibfnamefont {H.~G.}\ \bibnamefont {Craighead}}, \ and\ \bibinfo {author}
  {\bibfnamefont {J.~M.}\ \bibnamefont {Parpia}},\ }\bibfield  {title}
  {\enquote {\bibinfo {title} {Modal dependence of dissipation in silicon
  nitride drum resonators},}\ }\href {\doibase 10.1063/1.3671150} {\bibfield
  {journal} {\bibinfo  {journal} {Applied Physics Letters}\ }\textbf {\bibinfo
  {volume} {99}},\ \bibinfo {pages} {253103} (\bibinfo {year} {2011})},\
  \Eprint {http://arxiv.org/abs/https://doi.org/10.1063/1.3671150}
  {https://doi.org/10.1063/1.3671150} \BibitemShut {NoStop}%
\bibitem [{\citenamefont {Strauss}(1992)}]{strauss_PDE}%
  \BibitemOpen
  \bibfield  {author} {\bibinfo {author} {\bibfnamefont {W.~A.}\ \bibnamefont
  {Strauss}},\ }\href@noop {} {\emph {\bibinfo {title} {Partial differential
  equations - an introduction}}}\ (\bibinfo  {publisher} {John Wiley and Sons,
  Inc.},\ \bibinfo {year} {1992})\BibitemShut {NoStop}%
\bibitem [{\citenamefont {Poot}\ and\ \citenamefont {van~der
  Zant}(2012)}]{poot_physrep_quantum_regime}%
  \BibitemOpen
  \bibfield  {author} {\bibinfo {author} {\bibfnamefont {M.}~\bibnamefont
  {Poot}}\ and\ \bibinfo {author} {\bibfnamefont {H.~S.}\ \bibnamefont {van~der
  Zant}},\ }\bibfield  {title} {\enquote {\bibinfo {title} {Mechanical systems
  in the quantum regime},}\ }\href {\doibase 10.1016/j.physrep.2011.12.004}
  {\bibfield  {journal} {\bibinfo  {journal} {Phys. Rep.}\ }\textbf {\bibinfo
  {volume} {511}},\ \bibinfo {pages} {273--335} (\bibinfo {year}
  {2012})}\BibitemShut {NoStop}%
\bibitem [{\citenamefont {Lide}(1974)}]{lide_chemphys}%
  \BibitemOpen
  \bibinfo {editor} {\bibfnamefont {D.~R.}\ \bibnamefont {Lide}},\ ed.,\ \href
  {http://www.hbcpnetbase.com/} {\emph {\bibinfo {title} {Handbook of Chemistry
  and Physics}}}\ (\bibinfo  {publisher} {GRC press},\ \bibinfo {year}
  {1974})\BibitemShut {NoStop}%
\bibitem [{\citenamefont {Miller}\ and\ \citenamefont
  {Alemán}(2019)}]{miller_APL_mode_map_blue_red}%
  \BibitemOpen
  \bibfield  {author} {\bibinfo {author} {\bibfnamefont {D.}~\bibnamefont
  {Miller}}\ and\ \bibinfo {author} {\bibfnamefont {B.}~\bibnamefont
  {Alemán}},\ }\bibfield  {title} {\enquote {\bibinfo {title} {Spatially
  resolved optical excitation of mechanical modes in graphene nems},}\ }\href
  {\doibase 10.1063/1.5111755} {\bibfield  {journal} {\bibinfo  {journal}
  {Appl. Phys. Lett.}\ }\textbf {\bibinfo {volume} {115}},\ \bibinfo {pages}
  {193102} (\bibinfo {year} {2019})},\ \Eprint
  {http://arxiv.org/abs/https://doi.org/10.1063/1.5111755}
  {https://doi.org/10.1063/1.5111755} \BibitemShut {NoStop}%
\bibitem [{\citenamefont {Rohse}\ \emph {et~al.}(2020)\citenamefont {Rohse},
  \citenamefont {Butlewski}, \citenamefont {Klein}, \citenamefont {Wagner},
  \citenamefont {Friesen}, \citenamefont {Schwarz}, \citenamefont
  {Wiesendanger}, \citenamefont {Sengstock},\ and\ \citenamefont
  {Becker}}]{rohse_RSI_cavity_lock_optical_spring}%
  \BibitemOpen
  \bibfield  {author} {\bibinfo {author} {\bibfnamefont {P.}~\bibnamefont
  {Rohse}}, \bibinfo {author} {\bibfnamefont {J.}~\bibnamefont {Butlewski}},
  \bibinfo {author} {\bibfnamefont {F.}~\bibnamefont {Klein}}, \bibinfo
  {author} {\bibfnamefont {T.}~\bibnamefont {Wagner}}, \bibinfo {author}
  {\bibfnamefont {C.}~\bibnamefont {Friesen}}, \bibinfo {author} {\bibfnamefont
  {A.}~\bibnamefont {Schwarz}}, \bibinfo {author} {\bibfnamefont
  {R.}~\bibnamefont {Wiesendanger}}, \bibinfo {author} {\bibfnamefont
  {K.}~\bibnamefont {Sengstock}}, \ and\ \bibinfo {author} {\bibfnamefont
  {C.}~\bibnamefont {Becker}},\ }\bibfield  {title} {\enquote {\bibinfo {title}
  {A cavity optomechanical locking scheme based on the optical spring
  effect},}\ }\href {\doibase 10.1063/5.0010255} {\bibfield  {journal}
  {\bibinfo  {journal} {Review of Scientific Instruments}\ }\textbf {\bibinfo
  {volume} {91}},\ \bibinfo {pages} {103102} (\bibinfo {year} {2020})},\
  \Eprint {http://arxiv.org/abs/https://doi.org/10.1063/5.0010255}
  {https://doi.org/10.1063/5.0010255} \BibitemShut {NoStop}%
\bibitem [{Note1()}]{Note1}%
  \BibitemOpen
  \bibinfo {note} {Note, that it is also possible to perform this task using
  the digital signal processor in the lock-in amplifier \cite
  {poot_PRA_squeezing_feedback,poot_NJP_Yfeedback}, which can further improve
  the operation speed.}\BibitemShut {Stop}%
\bibitem [{\citenamefont {Poot}\ \emph {et~al.}(2011)\citenamefont {Poot},
  \citenamefont {Etaki}, \citenamefont {Yamaguchi},\ and\ \citenamefont
  {van~der Zant}}]{poot_APL_cooling}%
  \BibitemOpen
  \bibfield  {author} {\bibinfo {author} {\bibfnamefont {M.}~\bibnamefont
  {Poot}}, \bibinfo {author} {\bibfnamefont {S.}~\bibnamefont {Etaki}},
  \bibinfo {author} {\bibfnamefont {H.}~\bibnamefont {Yamaguchi}}, \ and\
  \bibinfo {author} {\bibfnamefont {H.~S.~J.}\ \bibnamefont {van~der Zant}},\
  }\bibfield  {title} {\enquote {\bibinfo {title} {Discrete-time quadrature
  feedback cooling of a radio-frequency mechanical resonator},}\ }\href
  {\doibase 10.1063/1.3608148} {\bibfield  {journal} {\bibinfo  {journal} {Appl
  Phys Lett}\ }\textbf {\bibinfo {volume} {99}},\ \bibinfo {eid} {013113}
  (\bibinfo {year} {2011})}\BibitemShut {NoStop}%
\bibitem [{\citenamefont {Cole}\ \emph {et~al.}(2011)\citenamefont {Cole},
  \citenamefont {Wilson-Rae}, \citenamefont {Werbach}, \citenamefont {Vanner},\
  and\ \citenamefont {Aspelmeyer}}]{cole_natcomm_phonon_tunnelling}%
  \BibitemOpen
  \bibfield  {author} {\bibinfo {author} {\bibfnamefont {G.~D.}\ \bibnamefont
  {Cole}}, \bibinfo {author} {\bibfnamefont {I.}~\bibnamefont {Wilson-Rae}},
  \bibinfo {author} {\bibfnamefont {K.}~\bibnamefont {Werbach}}, \bibinfo
  {author} {\bibfnamefont {M.~R.}\ \bibnamefont {Vanner}}, \ and\ \bibinfo
  {author} {\bibfnamefont {M.}~\bibnamefont {Aspelmeyer}},\ }\bibfield  {title}
  {\enquote {\bibinfo {title} {Phonon-tunnelling dissipation in mechanical
  resonators},}\ }\href {\doibase 10.1038/ncomms1212} {\bibfield  {journal}
  {\bibinfo  {journal} {Nature Communications}\ }\textbf {\bibinfo {volume}
  {2}},\ \bibinfo {pages} {231} (\bibinfo {year} {2011})}\BibitemShut {NoStop}%
\bibitem [{\citenamefont {Sun}\ \emph {et~al.}(2012)\citenamefont {Sun},
  \citenamefont {Zheng}, \citenamefont {Poot}, \citenamefont {Wong},\ and\
  \citenamefont {Tang}}]{sun_NL_double_beam}%
  \BibitemOpen
  \bibfield  {author} {\bibinfo {author} {\bibfnamefont {X.}~\bibnamefont
  {Sun}}, \bibinfo {author} {\bibfnamefont {J.}~\bibnamefont {Zheng}}, \bibinfo
  {author} {\bibfnamefont {M.}~\bibnamefont {Poot}}, \bibinfo {author}
  {\bibfnamefont {C.~W.}\ \bibnamefont {Wong}}, \ and\ \bibinfo {author}
  {\bibfnamefont {H.~X.}\ \bibnamefont {Tang}},\ }\bibfield  {title} {\enquote
  {\bibinfo {title} {Femtogram doubly clamped nanomechanical resonators
  embedded in a high-q two-dimensional photonic crystal nanocavity},}\ }\href
  {\doibase 10.1021/nl300142t} {\bibfield  {journal} {\bibinfo  {journal} {Nano
  Letters}\ }\textbf {\bibinfo {volume} {12}},\ \bibinfo {pages} {2299--2305}
  (\bibinfo {year} {2012})},\ \Eprint
  {http://arxiv.org/abs/http://pubs.acs.org/doi/pdf/10.1021/nl300142t}
  {http://pubs.acs.org/doi/pdf/10.1021/nl300142t} \BibitemShut {NoStop}%
\bibitem [{Note2()}]{Note2}%
  \BibitemOpen
  \bibinfo {note} {In principle, the response would be a sum over all modes
  $H_\protect \mathrm {total}(\omega ) = \DOTSB \sum@ \slimits@ _{m,n}
  H_{m,n}(\omega )$ but for simplicity it is assumed here that only a single
  mode $(m,n)$ gets excited. In Fig. \ref {fig:triplet}(a) and Fig. \ref
  {fig:31_and_13} the combined responses of respectively three and two modes
  are fitted to the data.}\BibitemShut {Stop}%
\bibitem [{Note3()}]{Note3}%
  \BibitemOpen
  \bibinfo {note} {Further refinements in the readout model could include the
  finite spot size of the laser by writing $\protect \ensuremath {\partial
  P/\partial U_{m,n}}$ as a two dimensional convolution of the beam shape and
  the mode shape.}\BibitemShut {Stop}%
\bibitem [{\citenamefont {Jhung}\ and\ \citenamefont
  {Jeong}(2015)}]{jhung_NET_vibrations_perforated_plates}%
  \BibitemOpen
  \bibfield  {author} {\bibinfo {author} {\bibfnamefont {M.~J.}\ \bibnamefont
  {Jhung}}\ and\ \bibinfo {author} {\bibfnamefont {K.~H.}\ \bibnamefont
  {Jeong}},\ }\bibfield  {title} {\enquote {\bibinfo {title} {Free vibration
  analysis of perforated plate with square penetration pattern using equivalent
  material properties},}\ }\href {\doibase
  https://doi.org/10.1016/j.net.2015.01.012} {\bibfield  {journal} {\bibinfo
  {journal} {Nuclear Engineering and Technology}\ }\textbf {\bibinfo {volume}
  {47}},\ \bibinfo {pages} {500--511} (\bibinfo {year} {2015})}\BibitemShut
  {NoStop}%
\bibitem [{\citenamefont {Poot}, \citenamefont {Fong},\ and\ \citenamefont
  {Tang}(2014)}]{poot_PRA_squeezing_feedback}%
  \BibitemOpen
  \bibfield  {author} {\bibinfo {author} {\bibfnamefont {M.}~\bibnamefont
  {Poot}}, \bibinfo {author} {\bibfnamefont {K.~Y.}\ \bibnamefont {Fong}}, \
  and\ \bibinfo {author} {\bibfnamefont {H.~X.}\ \bibnamefont {Tang}},\
  }\bibfield  {title} {\enquote {\bibinfo {title} {Classical non-gaussian state
  preparation through squeezing in an optoelectromechanical resonator},}\
  }\href {\doibase 10.1103/PhysRevA.90.063809} {\bibfield  {journal} {\bibinfo
  {journal} {Phys. Rev. A}\ }\textbf {\bibinfo {volume} {90}},\ \bibinfo
  {pages} {063809} (\bibinfo {year} {2014})}\BibitemShut {NoStop}%
\bibitem [{\citenamefont {Poot}, \citenamefont {Fong},\ and\ \citenamefont
  {Tang}(2015)}]{poot_NJP_Yfeedback}%
  \BibitemOpen
  \bibfield  {author} {\bibinfo {author} {\bibfnamefont {M.}~\bibnamefont
  {Poot}}, \bibinfo {author} {\bibfnamefont {K.~Y.}\ \bibnamefont {Fong}}, \
  and\ \bibinfo {author} {\bibfnamefont {H.~X.}\ \bibnamefont {Tang}},\
  }\bibfield  {title} {\enquote {\bibinfo {title} {Deep feedback-stabilized
  parametric squeezing in an opto-electromechanical system},}\ }\href {\doibase
  10.1088/1367-2630/17/4/043056} {\bibfield  {journal} {\bibinfo  {journal}
  {New J Phys}\ }\textbf {\bibinfo {volume} {17}},\ \bibinfo {pages} {043056}
  (\bibinfo {year} {2015})}\BibitemShut {NoStop}%
\end{thebibliography}%

\end{document}